\documentclass[12pt]{article}
\usepackage{graphicx}
\usepackage{cite}
\usepackage{amssymb,amsmath}
\begin{document}
\title{Possible scenarios of DNA double helix unzipping process}%
\author{Oleksii Zdorevskyi, Sergey N. Volkov\\
Bogolyubov Institute for Theoretical Physics, NAS of
Ukraine,\\14-b Metrolohichna Str., Kiev 03143, Ukraine \\
snvolkov@bitp.kiev.ua }\maketitle
\setcounter{page}{1}%
\maketitle
\begin{abstract}
Analysis of single-molecule micromanipulation experiments of DNA unzipping process shows some features of the force-distance curve, namely two consequent plateaus  in the area of ${\sim}12-14pN$ dependent on nucleotide sequence structure, as well as peaks appearance in the plateau area, to which it was not paid essential attention earlier. Using atom-atom potential function method the estimations of Watson-Crick base pairs opening energies are made. On the basis of this results two possible scenarios of the DNA double helix unzipping process are proposed. According to the first scenario DNA unzipping takes place slowly and as equilibrium process, with small difference between two plateaus on the unzipping curve. In this case firstly base pairs transit into the {\textquoteleft}pre-opened{\textquoteright} metastable state along the {\textquoteleft}opening{\textquoteright} pathway and then open along the {\textquoteleft}stretch{\textquoteright} pathway. Our estimations show that an important factor for the realization of this scenario is the existence of double-stranded DNA coil in the unopened part of DNA. The second scenario is characterized by higher opening force. In this scenario base pairs open directly along the {\textquoteleft}stretch{\textquoteright} pathway as non-equilibrium process. The conditions of the first scenario realization show that it can play a key role in the understanding of the DNA unzipping \textit{in vivo} during transcription and genetic information transfer processes.
\end{abstract}
\section{Introduction}
\label{intro}
The appearance of single-molecule micromanipulation methods~\cite{smith1992} made it possible to investigate biological macromolecules, considering it as a separate system. These methods use optical tweezers~\cite{simmons1996}, microneedles~\cite{3Bockelmann}, atomic force microscopes~\cite{bustamante1995}. Single-molecule micromanipulation methods allow to study processes such as DNA stretching, bending, twisting~\cite{lavery2002,bustamante2000, bustamante2003} and the DNA double helix nucleic base pairs opening~\cite{Destainville2016,1Bockelmann,2Bockelmann,3Bockelmann,4Bockelmann,5Bockelmann,6Bockelmann}. Among them  especially important is the process of sequential separation of nucleic bases in DNA base pairs under the action of external forces (DNA unzipping), which plays a key role in the genetic information transferring process.

At the early studies separation of bases in DNA double helix nucleic pairs ($bp$) was identified as the process of denaturation of macromolecules under the temperature influence. This process is called DNA melting~\cite{plrevmfk,owczarzy2004, wartell1985PhReport}. In this case the process of base pairs opening is not sequential and the denaturation takes place in DNA sites which have  different size and location. Firstly base pairs in A{\textperiodcentered}T-rich sites open, and then - in G{\textperiodcentered}C-rich sites. In contrast to the melting process, single-molecule micromanipulation methods make it possible to investigate the process of sequential opening of DNA base pairs in the same way as it takes place in cell \emph{in vivo}.

In the physical experiment of the unzipping process, which is carried out by the
single-molecule micromanipulation methods, the dependence of the
opening force on the opening distance is measured~\cite{1Bockelmann,2Bockelmann,3Bockelmann,4Bockelmann,5Bockelmann,6Bockelmann,7Bockelmann,ritort2010}. For $\lambda$-phage DNA  at small pulling velocities (${\sim}40nm/s$) the strand separation starts to take place after reaching by the force of a some critical value (${\sim}13pN$). For this force value two consequent plateaus at ${\sim}13 pN$ and ${\sim}12 pN$ are observed and force peaks ${\sim}1 pN$ in the plateau area occur ({\textquoteleft}sawtooth{\textquoteright} nature of the DNA unzipping process)~\cite{1Bockelmann}. For big velocities (${\geq}1 \mu m/sec$) the plateau has a bigger value and hysteresis is observed~\cite{4Bockelmann}.

The DNA unzipping process has been studied theoretically~\cite{3Bockelmann,2Bockelmann,ritort2010,voulgarakis2006,volkovUnz2009}. By Bockelmann et al.~\cite{2Bockelmann,3Bockelmann} a model of the unzipping process which takes into consideration the opening energies of base pairs and the elastic energies of single strands is proposed. The authors managed to obtain the critical force value of ${\sim}13pN$. The peaks occurrence in the plateau area is considered as a molecular {\textquoteleft}stick-slip motion{\textquoteright} that takes place because of a heterogeneity of macromolecule. But this model does not describe the difference between two plateaus on the $\lambda$-DNA unzipping curve. Probably, the mechanism of the unzipping process can differ from that described in the works~\cite{2Bockelmann,3Bockelmann}. In the present work it will be shown that the reason of its difference can be the base pair opening along different pathways.

The different pathways of base pairs opening are studied in the paper~\cite{volkovUnz2009} where two-component model describing the DNA unzipping process is proposed. According to this model the DNA unzipping process takes place due to the creation of conformational bistability in the system under the external force action. As a result the process of the base pairs opening flows cooperatively.

The goal of the present work is to determine the physical processes that occur in DNA macromolecule during the DNA unzipping and to describe the experimental data on the qualitative and quantitative level. In Sec.\ref{exp_an} the data from the single-molecule micromanipulation DNA unzipping experiment is analyzed and some features of the opening force-distance dependence that have not been explained in the previous theoretical models are determined. In Sec.\ref{main} using the atom-atom potential functions method A{\textperiodcentered}T and G{\textperiodcentered}C base pairs separation energies along different pathways are estimated. As the result it is determined that during the unzipping process DNA base pairs can open not only along the
{\textquoteleft}stretch{\textquoteright} pathway, but under some conditions they can firstly transit into a metastable state along the {\textquoteleft}opening{\textquoteright} pathway, and then fully
open along the {\textquoteleft}stretch{\textquoteright} pathway.

\section{Experimental data analysis}
\label{exp_an}
\begin{figure}
\begin{center}
\resizebox{0.7\textwidth}{!}{%
  \includegraphics{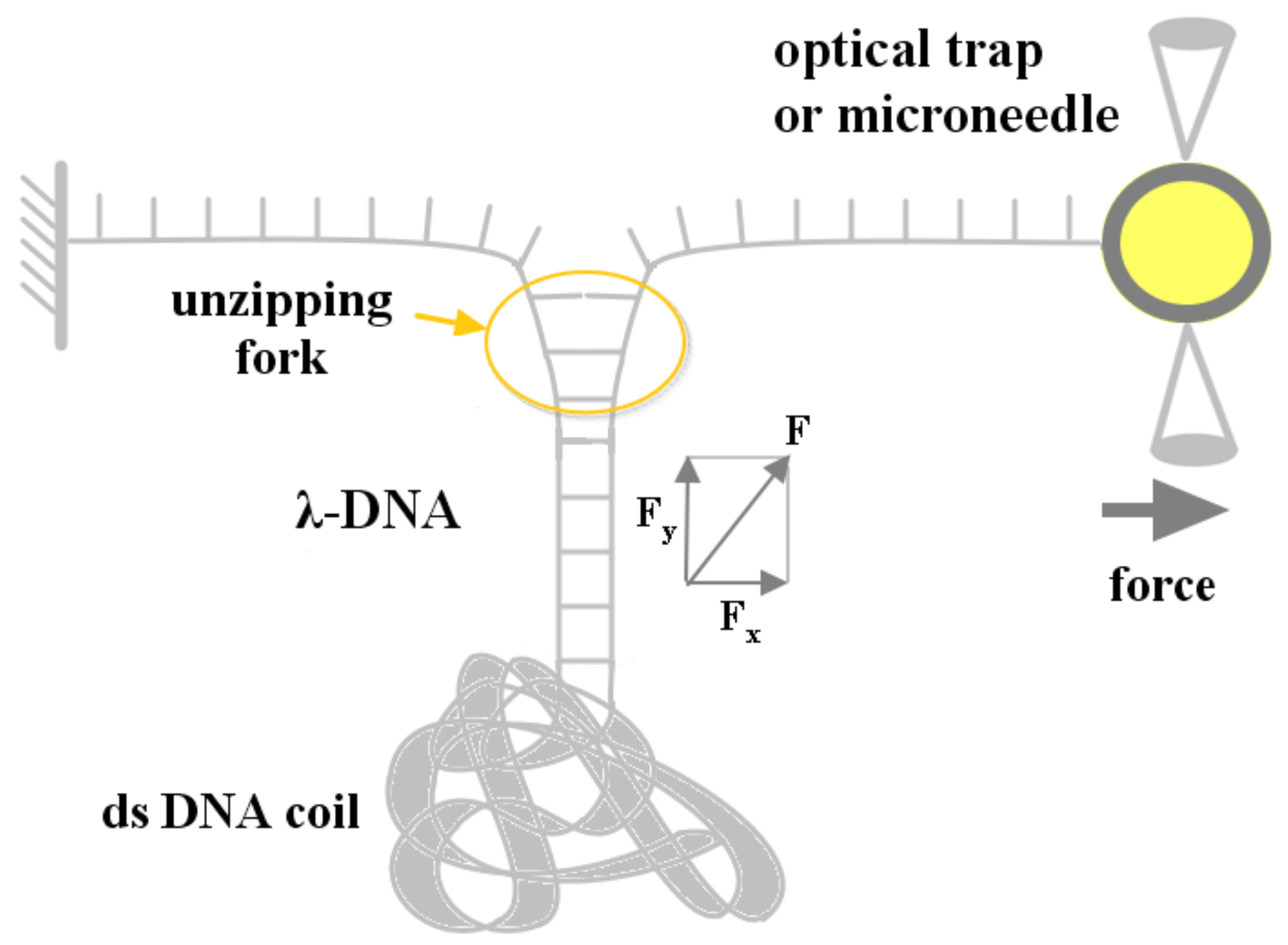}
}
\caption{Scheme of the single-molecule micromanipulation DNA unzipping
experiment~\cite{1Bockelmann,2Bockelmann,3Bockelmann,4Bockelmann,5Bockelmann,6Bockelmann,7Bockelmann}. $F$ - the force that acts on the area situated between the double-stranded DNA (dsDNA) coil and unzipping fork and its components $F_x$ and $F_y$.  }
\label{fig:expSheme}       
\end{center}
\end{figure}

In the single-molecule micromanipulation experiments the $\lambda
$-phage DNA double helix is studied~\cite{1Bockelmann,2Bockelmann,3Bockelmann,4Bockelmann,5Bockelmann,6Bockelmann,7Bockelmann}. It has a length of 48 502 bp. One end of one strand is
attached to the substrate, and another strand from the same end - to the ball,
which is held with a glass microneedle or optical trap. The scheme of the experiment is shown on Fig.\ref{fig:expSheme}. From the other end double helix is not fixed and it forms a tertiary
structure in the shape of a coil. Ball is moved with a constant speed,
thus consequently opening the double helix. Than the dependence of the
applied force on the displacement of the ball is measured.

\begin{figure}
\begin{center}
\resizebox{1\textwidth}{!}{%
  \includegraphics{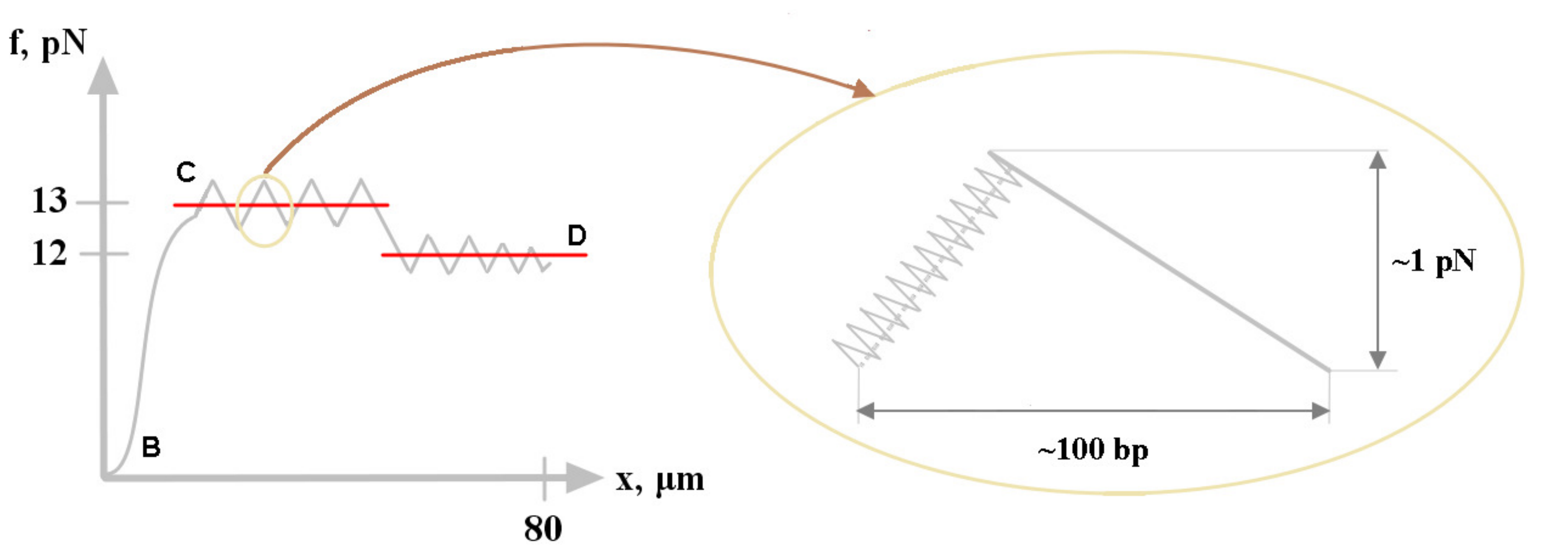}
}
\caption{Schematic figure of the experimental dependence obtained in~\cite{1Bockelmann}. Our estimations reveal that two plateaus have a value of ${\sim}12 pN$ and ${\sim}13pN$ and peaks have the height of ${\sim}1 pN$ and the length of
${\sim}100 bp$; small peaks on the ascents of big peaks.}
\label{fig:expDepen}       
\end{center}
\end{figure}

Let us analyse the curve structure (Fig.\ref{fig:expDepen}) obtained in
the~\cite[fig.3]{1Bockelmann}. On the curve fragment B - C sharp increase of
the force almost without changing of the distance is observed. This
corresponds to the tension increase of the whole system until
force reaches a certain critical value, which corresponds to the bases separation in DNA nucleic pairs.

On the curve fragment C - D base pairs opening takes place. This fragment consists of two plateaus.
In the first half of the double helix opening occurs on the average force value of ${\sim}13 pN$, and in the second half on ${\sim}12 pN $ (Fig.\ref{fig:expDepen}). As seen, there are two different plateaus with a small difference in the critical force value.

It is known that the first half of $\lambda $-DNA consists mainly of \ G{\textperiodcentered}C pairs, while the second half is dominated by A{\textperiodcentered}T pairs. Therefore, the average opening force of the first half of $\lambda$-DNA will be close to the opening force of the G{\textperiodcentered}C pair, and in the second half - to the opening force of the A{\textperiodcentered}T pair. So, the unzipping of these two parts of $\lambda$-phage DNA should take place under different opening force. Really, as it is seen from the experiment the value of force difference is ${\sim}1 pN$ (Fig.\ref{fig:expDepen}).

\begin{table}
\begin{center}
\noindent\caption{Free opening energies of A{\textperiodcentered}T and G{\textperiodcentered}C base pairs, calculated in the corresponding works, the difference between the opening forces of A{\textperiodcentered}T and G{\textperiodcentered}C base pairs  $\Delta F=F_{GC}-F_{AT} $ and the averange opening force $F_{av}=(F_{GC} +F_{AT})/2 $. $^*$fitting parameters in~\cite{3Bockelmann}, $^{**}$opening  energies of  A{\textperiodcentered}T- and G{\textperiodcentered}C- containing polymers, averaged over the nearest neighbors~\cite{mfk2006}, $^{***}$our estimations using (\ref{for:op_force}), $^{****}$calculated in~\cite{3Bockelmann}.}\vskip3mm\tabcolsep4.5pt
\label{tab:force_exp}
\noindent{\footnotesize\begin{tabular}{|c|c|c|c|c|}
\hline
 &~\cite{3Bockelmann}&~\cite{Breslauer1986}&~\cite{mfk2006}&~\cite{ritort2010}  \\
 \hline
Conditions (t, {\textordmasculine}C;NaCl, mM) & 25{\textordmasculine}C, 150mM & 25{\textordmasculine}C,10\textsuperscript{3} mM & 37{\textordmasculine}C, 15 mM & 25{\textordmasculine}C, 10\textsuperscript{3} mM \\
 \hline
G\textsubscript{AT}, kcal/mol &
{}-0.77$^*$ & {}-1.55$^{**}$ & {}-0.31$^{**}$ &
{}-1.09$^{**}$\\
\hline
G\textsubscript{GC}, kcal/mol & {}-1.7$^*$ & {}-3.23$^{**}$ &
{}-1.45$^{**}$ & {}-2.08$^{**}$ \\
\hline
${\Delta F, pN}^{***}$ &  7 &  12 & 8 &  7\\
\hline
F\textsubscript{av}, pN & 13$^{****}$ & 22$^{***}$ & 11$^{***}$ &
16$^{***}$\\
\hline
\end{tabular}}
\end{center}
\end{table}

In the paper~\cite{3Bockelmann} a model for the DNA unzipping process is proposed. In the framework of this model, the force of the base pairs separation in the unzipping process of DNA double helix can be calculated as:
\begin{equation}
\label{for:op_force}
 F = \dfrac{E_{pair}+2E_{el}}{l}  \ ,
\end{equation}
 where  ${E}_{\mathit{pair}}$ is the the total energy of base pairs opening in the macromolecule,  ${E}_{\mathit{el}}$ is the extension energy of one single strand, $l$ - total extension of the single strands.
From freely-joined chain model: ${E}_{\mathit{el}}{\approx}0.3\mathit{kcal}/{\mathit{mol}},l=0.95\mathit{nm}$~\cite{3Bockelmann}. In the paper~\cite{3Bockelmann} fitting parameters for A{\textperiodcentered}T and G{\textperiodcentered}C opening energies are used (see Tabl.\ref{tab:force_exp}) and the average opening force of a polymer with 50\% content of A{\textperiodcentered}T and G{\textperiodcentered}C pairs is estimated. It coincides with the experimental value ($\Delta $F, Tabl.\ref{tab:force_exp}). 

However, if we estimate by this formula the opening forces of one pair in A{\textperiodcentered}T- and G{\textperiodcentered}C-containing polymers separately ($F_{AT}$ and $F_{GC}$ respectively), and calculate their difference $\Delta F =
F_{GC}-F_{AT}$, we will get a value which is large enough and does not coincide with the experimental difference between two plateaus of the value of ${\sim}1 pN $ (see. Tabl.\ref{tab:force_exp}). If as ${E}_{\mathit{pair}}$ parameters in formula (\ref{for:op_force}) we take base pairs opening energies obtained from the DNA melting experiments~\cite{Breslauer1986, mfk2006, ritort2010} and then average them over all nearest neighbors, we will get a value of $\Delta F$, which is ${\sim}7-12 pN$ and does not match the experimental data. 

Note, in the works~\cite{Breslauer1986, ritort2010} the average opening force values $ F_{av}=(F_{AT}+F_{GC})/2$ are inflated as well. As energy data calculated from the results of the DNA melting experiment, using of these energies may be incorrect for the unzipping process. Also experimental studying or short A{\textperiodcentered}T and G{\textperiodcentered}C DNA hairpins shows the difference $\Delta F{\approx}10pN$~\cite{Destainville2016,rief1999} which is also close to the $\Delta F$, obtained from the melting process studying (Tabl.\ref{tab:force_exp}). These facts can reveal that the base pair opening in the processes of DNA melting and unzipping of short DNA hairpins can take place along the one pathway but in the process of $\lambda$-phage DNA unzipping at low pulling velocity along the another pathway.

Let us consider the plateau region in more detail. There are big peaks with the height ${\sim}1 pN $ and with the length ${\sim}100bp$ up to $1000 bp$ (Fig.\ref{fig:expDepen}). It should be noted that the height of the peaks reduces significantly to the end of the unzipping process.

On the ascent of these peaks even smaller peaks ${\sim}10bp$ and
up to $0.1 pN$ are observed (Fig.\ref{fig:expDepen},~\cite[fig.6]{1Bockelmann}). The values related to the size of small peaks are approximate because of the lack of resolution of available data. Shape of small peaks is characterized by a negative slope, as if a ball which is pulled, occasionally starts to move back. The negative slope can correspond to the position fluctuations of the ball in the optical trap. In the paper~\cite{ritort2010} frequency filters and short linkers that connect the DNA strands with ball and substrate are used allowing to avoid small peaks. On the descents of big peaks there are no small peaks. This means that the ball position fluctuations on ascends of big peaks arising occur because of the growth of tension in the system and descents of big peaks are followed by a decrease in tension, so there are no fluctuations on the descents of big peaks. This shows that the process of base pairs opening flows cooperatively on the descends of big peaks. As the small peaks are not related to the physical mechanism of the DNA unzipping they will not be discussed in the present paper any more,  so {\textquoteleft}big peaks{\textquoteright} will be called simply {\textquoteleft}peaks{\textquoteright}.

Shape of the peaks presented on Fig.\ref{fig:expDepen} is uniform and periodic, however,
due to the heterogeneity and  aperiodicity of the coil-like tertiary structure of the $\lambda$-DNA double helix, the height and length of these peaks may be different from one to the other.

In the paper~\cite[p.548]{ritort2006} it is noted that when force is less than $10 pN$, opening of base pairs may not occur at all, but when force is {\textgreater} $17 pN$, base pair opening takes place very quickly and peaks on the obtained force-distance dependence reflect nucleotide sequence of the double helix. These opening force values can be obtained by increasing of the pulling speed~\cite{2Bockelmann,3Bockelmann,4Bockelmann,5Bockelmann,6Bockelmann}. At speeds greater than $1\mu m/s $ unzipping process becomes to be non-equilibrium and a hysteresis is observed~\cite{4Bockelmann}.

The process of the nucleic base pairs opening can be considered as a process of translational motion of the unzipping fork along the double helix~\cite{2Bockelmann}. The progressive movement of the fork is not uniform, and the size of the fork changes all the time. Using the statistical mechanics method~\cite{3Bockelmann} it is shown that unzipping fork is occasionally blocked and this is the reason of the force growth  on the peaks. After that a cooperative process of base pairs opening takes place which corresponds to the forward movement of the unzipping fork and the descent of the peak. Our estimates (Fig.\ref{fig:expDepen}) show that the probable length of the cooperative opening of DNA sites is in the range of several hundred pairs. At the same time in~\cite{3Bockelmann} the possible mechanism of the fork blocking and, as a result of the peak occurrence, is considered to be the heterogeneity of the macromolecule. But it should be noted, that heterogeneity can manifest itself in scale of
several base pairs and can not be significantly noticeable on such big scales as several hundreds base pairs. As it will be shown in Sec.\ref{main}, the presence of peaks may occur due to the coil-like tertiary structure of the closed part of DNA double helix in solution.

Thus, our analysis shows some important features of the DNA unzipping process, such as a difference between the critical opening forces for the two parts of $\lambda$-phage DNA, the presence of peaks and their shape characterize the unzipping process. These features require detailed physical explanation to understand all of the stages of the DNA unzipping process.

\section{ Scenario of the DNA unzipping process. Some estimations}
\label{main}
\subsection{Mechanism of the peaks occurrence on the unzipping curve. }
\label{peaks}
Let us consider the detailed shape of the force-distance curve  in the plateau area. This section will show how to explain a scenario of the DNA unzipping process using the experimental results shown on Fig.\ref{fig:expDepen}.

\begin{figure}
\begin{center}
\resizebox{0.4\textwidth}{!}{%
  \includegraphics{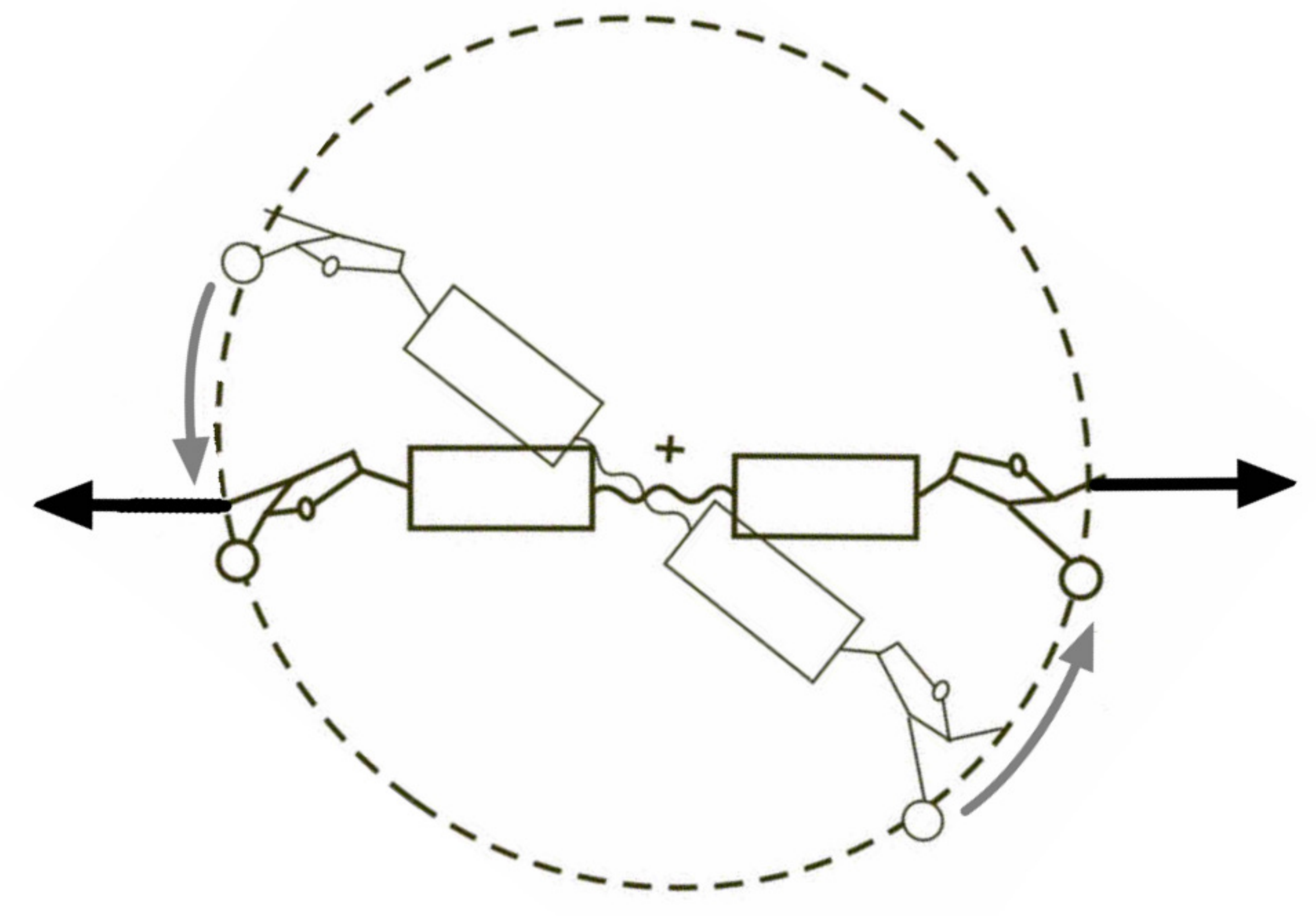}
}
\caption{Base pair twist --- the reason of
the DNA double helix rotation during unzipping process.}
\label{fig:twist}       
\end{center}
\end{figure}

As the plateau consists of peaks, to understand the unzipping process scenario, let us consider the mechanism of peaks occurrence in detail. As it is known from the early study~\cite{levinthal1956}, the DNA unzipping process is accompanied by rotation of the double helix around its axis. Molecular dynamics methods~\cite{VolkovMolDynSym} showed that this rotation happens due to the fact that the projection of each subsequent couple on a plane perpendicular to the axis of the double helix, is turned on an angle with respect to the previous (Fig.\ref{fig:twist}). As it was already mentioned the unopened part of the DNA double helix has a coil-like shape in solution (Fig.\ref{fig:expSheme}). So according to this scenario this coil must rotate during the unzipping process. At the beginning the tension that
occurs in the DNA site situated	 between the fork and the coil is not enough to rotate the coil, but when the force reaches some threshold value, the coil is rotated and the tension disappears. From simple
hydrodynamic considerations we can determine on which parameters the force required to rotate the coil at a certain angular velocity depends. The radius of the coil is given by:

\begin{equation}
R=\sqrt{2L{l}_{p}} \ ,
\end{equation}

where $L$ is the contour length, and  ${l}_{p}=50\mathit{nm}$ is the persistent
length of DNA~\cite{Saenger}.

For simplicity we consider the coil as a smooth ball. Torque that should
be applied to the ball of radius $R$ in order to rotate it in the medium
of a dynamic viscosity coefficient $\mu$ with the angular velocity
$\omega$, is as given by~\cite{slezkin}:
\begin{equation}
\left|M\right|=8\pi {\ast}9.81\mathit{\mu \omega }{R}^{3} \ .
\end{equation}
As $M=\mathit{FR}$,
\begin{equation}
\label{for:proportion}
F{\sim}{R}^{2}{\sim}L \ .
\end{equation}

Since the contour length of the double helix decreases with the DNA
opening, so this force, and therefore the peak height should decrease.
This decrease in peak height is currently observed on the unzipping
curve. While the peak height depends on the force of a coil rotation, the
peak length is defined by other parameters. On the peak ascent
unzipping fork is blocked~\cite{3Bockelmann}, so the ascent length is mostly
determined by the extension of single strands (Fig.\ref{fig:expSheme}). The length of
the descent is determined by the number of opened pairs.

The presence of peaks is also observed in the unpeeling process, when
the force is applied at one end to one of the strands of the double
helix, and at the other end both strands are fixed~\cite{gross2011unpeeling}. However, in the force-distance curve there are significant differences from the unzipping process. Firstly, the peak
height increases at the end of the opening of $\lambda$-DNA. Secondly,
compared with the unzipping process the height of peaks is much
bigger. These features may also be connected with the coil mechanics.
Unlike the unzipping process, coil size increases with  opening of the
double helix, so the height of peaks increases with the DNA
opening. Significantly bigger peaks height (order of magnitude more than
in the unzipping process) may occur due to the fact that in this
process a coil rotates not around its own axis but around the axis of
the helix, that significantly increases the hydrodynamic friction
force.

These facts indicate that the presence of peaks on the force-distance
curve of the DNA unzipping process can be related with the
hydrodynamics of a coil-like tertiary structure of the DNA double helix
in solution. Since a force required for rotating of this coil is proportional
to the contour length of DNA in a coil (\ref{for:proportion}) and in the experiment~\cite{1Bockelmann} (Fig.\ref{fig:expDepen}) the long(${\approx}50000bp)$ $\lambda$-DNA is used for
studying, this coil can play a significant role in the process of DNA
unzipping. So coil mechanics can be the main reason of unzipping fork
blocking mentioned in~\cite{3Bockelmann} and consequently of the peak occurence on the force-distance curve (Fig.\ref{fig:expDepen}).

\subsection{The possible pathways of the DNA unzipping process}
\label{scenario}
\begin{figure}
\begin{center}
\resizebox{0.5\textwidth}{!}{%
  \includegraphics{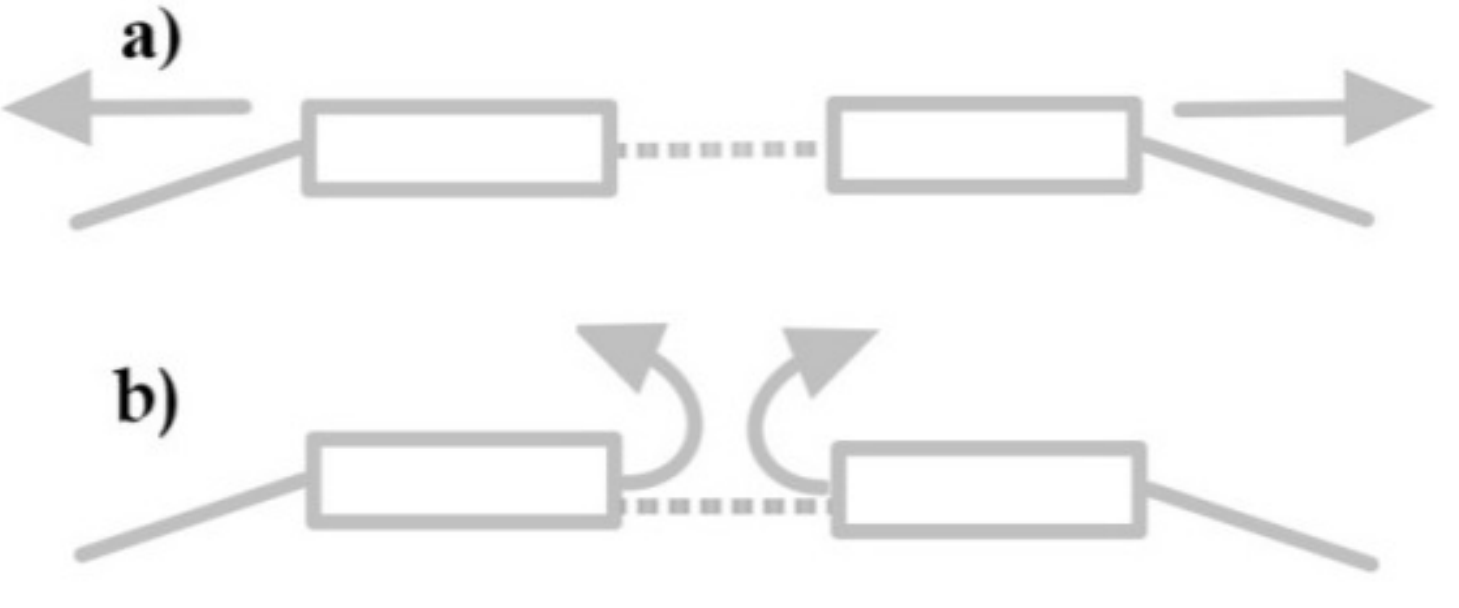}
}
\caption{Degrees of freedom of nucleic bases in a pair of nucleotides
(both movements take place in the plane of a pair): a) {\textquoteleft}stretch{\textquoteright} pathway
b){\textquoteleft}opening{\textquoteright} pathway.}
\label{fig:degFree}       
\end{center}
\end{figure}

Opening of nucleic base pairs can take place along different pathways~\cite{nomenclatureEMBO}. Among them the most probable for the DNA unzipping
process~\cite{volkovUnz2009,VolkovMolDynSym} are the
{\textquoteleft}stretch{\textquoteright} pathway, when bases in a
pair separate from each other along hydrogen bonds (Fig.\ref{fig:degFree}a), and the
{\textquoteleft}opening{\textquoteright} pathway, when base pairs open
into a groove (Fig.\ref{fig:degFree}b). Here all motions take place in the plane of a
pair~\cite{VolkovMolDynSym}.

\begin{figure}
\begin{center}
\resizebox{0.8\textwidth}{!}{%
  \includegraphics{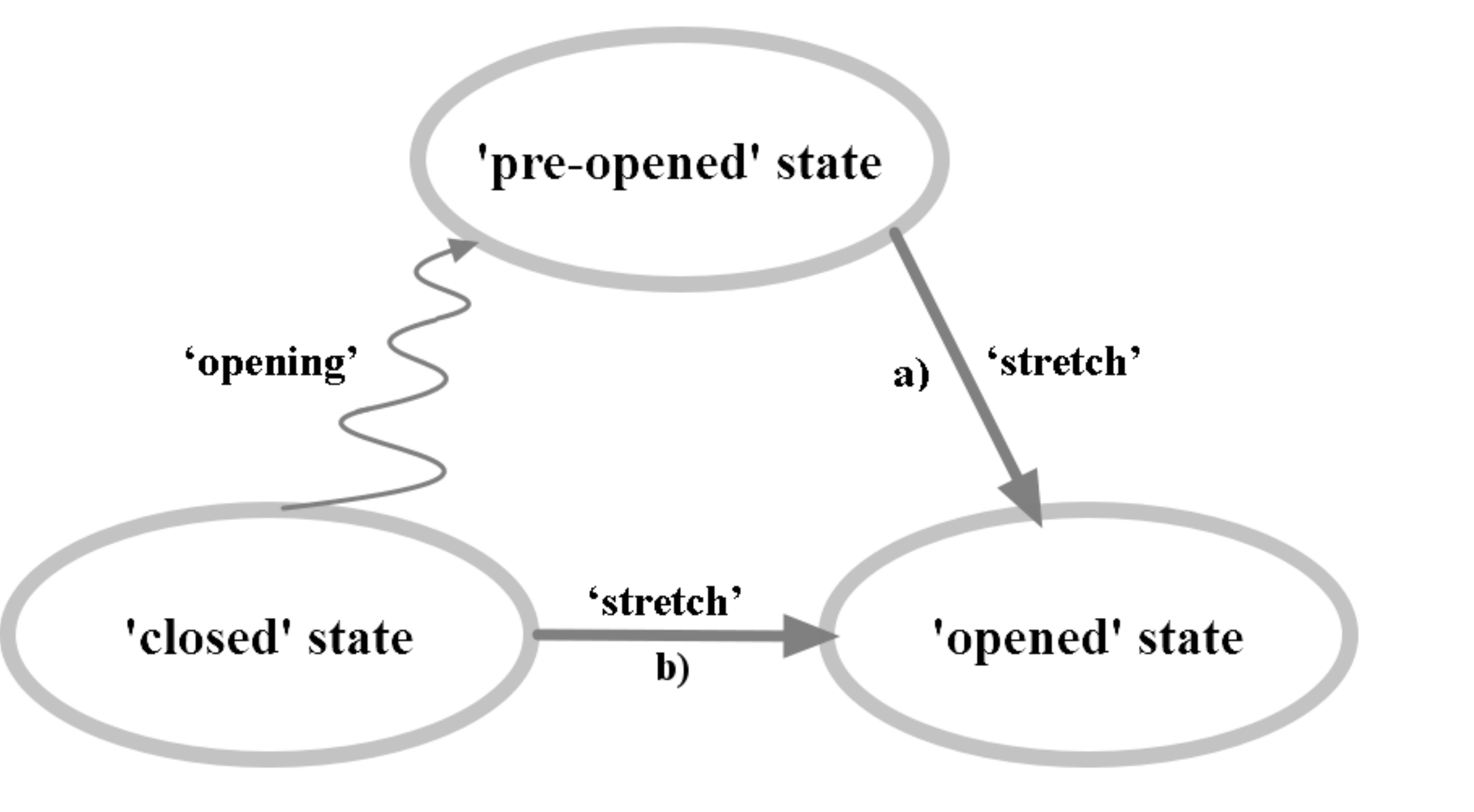}
}
\caption{Two possible scenarios of nucleic base pairs
separation in DNA unzipping process:
a) separation along the {\textquoteleft}stretch{\textquoteright}
pathway from the state which is pre-opened along the
{\textquoteleft}opening{\textquoteright} pathway; b) direct
separation along the {\textquoteleft}stretch{\textquoteright}
\ pathway.}
\label{fig:scenario}       
\end{center}
\end{figure}

Let us consider that the process of base pairs opening may take place according to two
scenarios: Watson-Crick pair opens along the {\textquoteleft}stretch{\textquoteright} pathway (Fig.\ref{fig:scenario}b) or Watson-Crick pair firstly transits into {\textquoteleft}pre-opened{\textquoteright} metastable state along the {\textquoteleft}opening{\textquoteright} pathway, and then fully opens along the {\textquoteleft}stretch{\textquoteright} pathway (Fig.\ref{fig:scenario}a).

As it was shown in Sec.\ref{peaks}, this coil can play a
significant role in the process of $\lambda $-DNA unzipping. Base pair
transition into a metastable state can occur in the area
between the unzipping fork and a coil under the tension that creates external
force from the one side and a coil from the other ($F_y$ component on Fig.\ref{fig:expSheme}). Under the influence of this tension double helix is stretched in this area, resulting in
the weakening of stacking interactions between adjacent pairs, which
mostly determines the stability of the double helix~\cite{mfk2006}. This
allows base pairs to transit into the
{\textquoteleft}pre-opened{\textquoteright} metastable state along the
{\textquoteleft}opening{\textquoteright} pathway. This conformational transition in the area situated between the unzipping fork and a coil is natural due to the geometry of our system. So we consider that this transition does not require much energy and estimate the opening energy from the {\textquoteleft}pre-opened{\textquoteright} to the {\textquoteleft}opened{\textquoteright} state in this scenario.  The existence of such a {\textquoteleft}pre-opened{\textquoteright} metastable state is shown in the works~\cite{VolkovKryachko, volkovUnz2009, VolkovSolovyov2007, VolkovMolBiol1995}.

\begin{figure}
\begin{center}
\resizebox{0.6\textwidth}{!}{%
  \includegraphics{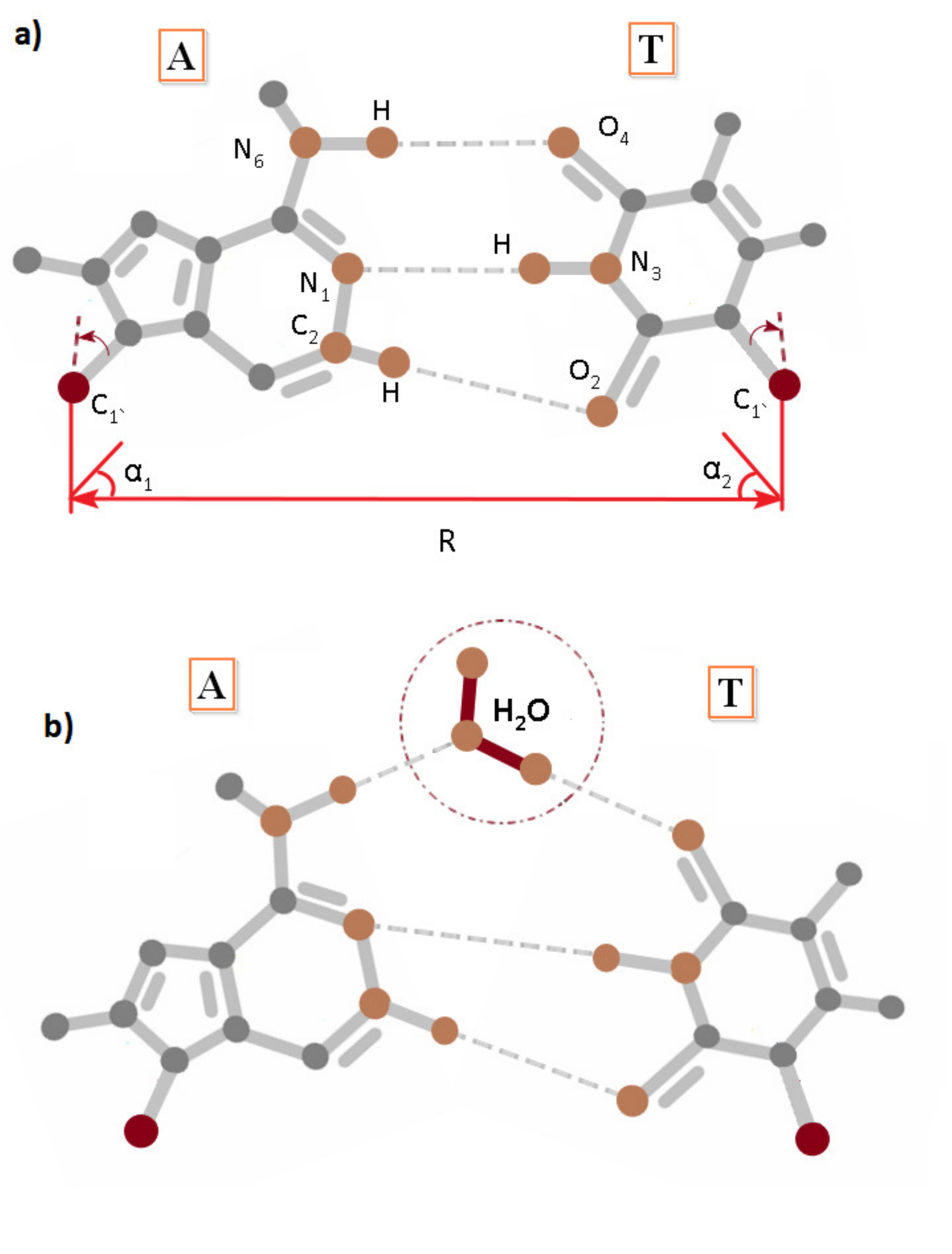}
}
\caption{Schematic representation of the
adenine-thymine nucleic base pair  in DNA macromolecule: a) Watson-Crick configuration~\cite{Saenger}. The arrow shows the {\textquoteleft}opening{\textquoteright} pathway: the bases
rotation around the axis passing through atoms $C_{1}'$
perpendicular to the plane of the pair; b) {\textquoteleft}pre-opened{\textquoteright}  A{\textperiodcentered}T pair mediated by the water molecule~\cite{VolkovKryachko}.}
\label{fig:ATpair}       
\end{center}
\end{figure}

Let us make some qualitative estimations of the energy required for base pair transition from Watson-Crick state to the opened state for A{\textperiodcentered}T and G{\textperiodcentered}C base pairs. In our estimations we consider that in the {\textquoteleft}opening{\textquoteright} pathway the rotation of bases takes place around the axis passing through the atom $C_{1}'$  perpendicular to the plane of the base (Fig.\ref{fig:ATpair}). As
the {\textquoteleft}stretch{\textquoteright} pathway we consider a
motion of bases in a pair relative to each other along the line
connecting the $C_{1}'$ atoms (Fig.\ref{fig:ATpair}).

Let us define the parameters of these processes. We consider that the state
becomes to be {\textquoteleft}pre-opened{\textquoteright} when the
external hydrogen bond of a pair is broken ($N_6H$
{\textbullet}{\textbullet}{\textbullet} $O_4$ for A{\textperiodcentered}T and $O_6$
{\textbullet}{\textbullet}{\textbullet} $HN_4$ to G{\textperiodcentered}C), i.e. the distance
between the heavy atoms ($N$ or $O$) and hydrogen atom reaches the value of
the sum of their van der Waals radii. As the van der Waals radius of hydrogen atom is ${\approx}1.2 ${\AA}, nitrogen and oxygen are
${\approx}1.5${\AA}~\cite{zefirov1974}, this distance will be ${\approx}2.7${\AA}. Taking
into account that the length of the covalent bond N-H ${\approx}1${\AA}~\cite{Saenger}, hydrogen bond will be broken when the distance between
heavy atoms reaches ${\approx}3.7${\AA}. We consider that the full
base pair opening takes place when the middle hydrogen bond ($N_1$
{\textbullet}{\textbullet}{\textbullet} $N_3$) is broken~\cite{plrevmfk}, i.e. the
distance $N_1N_3{\approx}3.7${\AA}. The estimations made in
this section take into account only one degree of freedom: rotation of
bases in a pair ({\textquoteleft}opening{\textquoteright} pathway) or movement
of bases relative to each other along the hydrogen bonds
({\textquoteleft}stretch{\textquoteright} pathway). However, in the
experiment the movement along this pathways may take place involving
other degrees of freedom ({\textquoteleft}twist{\textquoteright},
{\textquoteleft}propeller{\textquoteright},
{\textquoteleft}buckle{\textquoteright}, etc.~\cite{nomenclatureEMBO}),
which are not considered here to simplify the calculations because our aim is only to
give the understanding of the physical mechanisms which take
place in the DNA unzipping process.

To estimate the hydrogen bond stretching energy in Watson-Crick base pairs we use atom-atom potential functions method. Here we consider the pairwise interactions between the atoms which form hydrogen bonds.  The energy of the hydrogen bond between atoms i and j is modeled by modificated Lennard-Jones potential~\cite{poltevShul1984}:
\begin{equation}
\label{for:lenJones}
U\left({r}_{\mathit{ij}}\right)=\frac{-{A}_{\mathit{ij}}}{{r}_{ij}^{10}}+\frac{{B}_{\mathit{ij}}}{{r}_{ij}^{12}}
\end{equation}
Parameters $A_{ij}$ and $B_{ij}$ are also taken from~\cite{poltevShul1984}, and for $H$ {\textbullet}{\textbullet}{\textbullet} $N$ are $A_{HN} = 9100$ $kcal*${\AA}$^{10} / mol$, $B_{HN} = 27400$  $kcal*${\AA}$^{12} / mol$, and for bond $H$ {\textbullet}{\textbullet}{\textbullet} $O$ $A_{HO} = 7350$ $kcal*${\AA}$^{10}/mol$, $B_{HO} = 21400$ $kcal*${\AA}$^{12}/mol$. Geometry of Watson-Crick pairs A{\textperiodcentered}T and G{\textperiodcentered}C is taken from~\cite{Saenger}. For
the A{\textperiodcentered}T pair R distance between atoms $C_{1}'$ \ is taken
$10.44$ {\AA}, the angles between the glycosidic bond and R ($\alpha_1 $ and
$\alpha_2 $) are calculated from the condition that Watson-Crick A{\textperiodcentered}T
pair in the closed state has the distance $N_6H$
{\textbullet}{\textbullet}{\textbullet} $O_4$ of the value $2.9$ {\AA}, and the distance $N_1$
{\textbullet}{\textbullet}{\textbullet} $HN_3$ is $2.8$ {\AA}. For the pair G{\textperiodcentered}C
estimates are made in the similar way (see. Tabl.\ref{tab:param}). For A{\textperiodcentered}T pair the
shortened contact $C_2H$ {\textbullet}{\textbullet}{\textbullet} $O_2$ is
taken into account, which is also simulated by the potential (\ref{for:lenJones}).

\begin{figure}
\begin{center}
\resizebox{0.8\textwidth}{!}{%
  \includegraphics{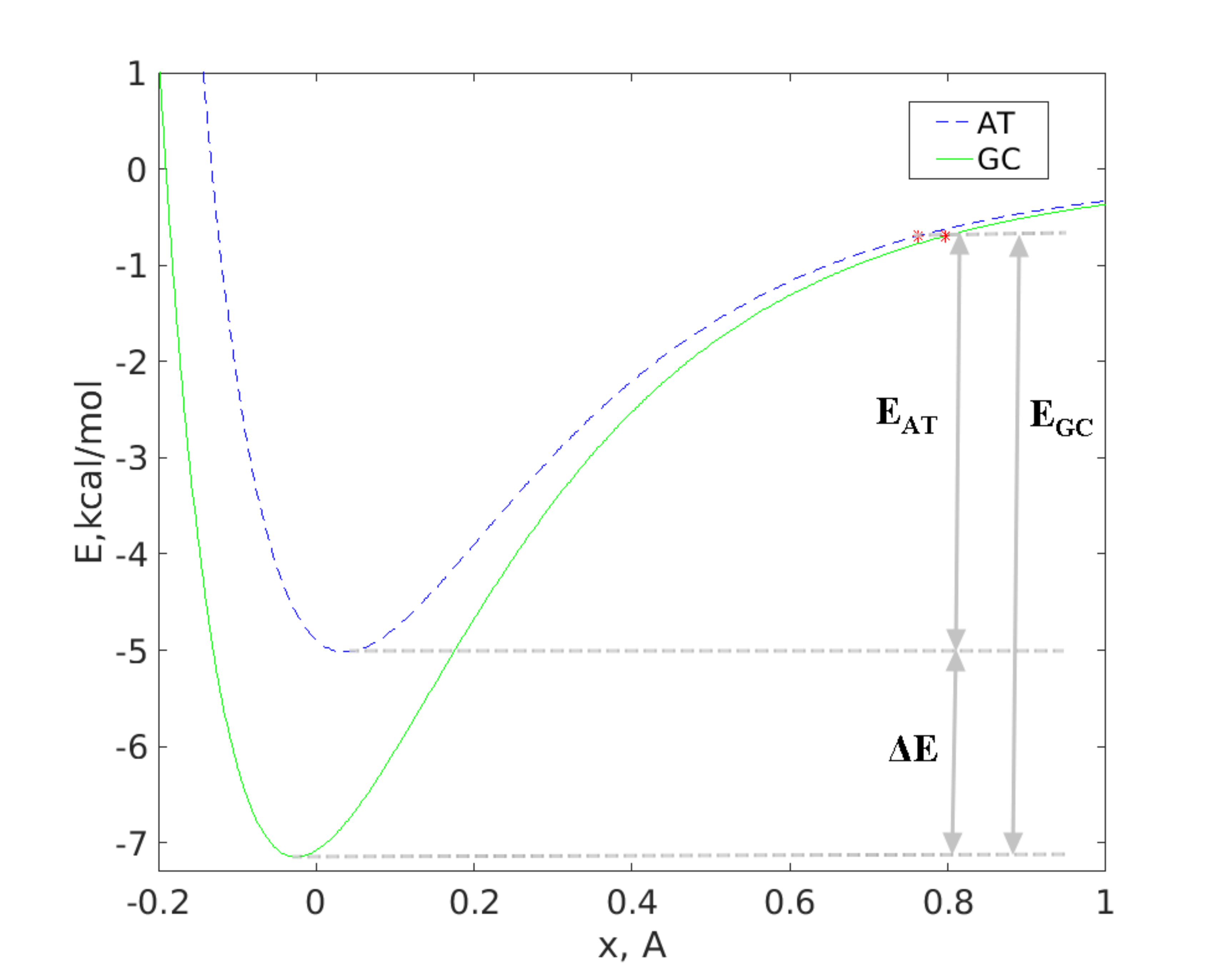}
}
\caption{Binding energy dependencies from distance calculated for A{\textperiodcentered}T (dashed line) and G{\textperiodcentered}C (solid line) base pairs for {\textquoteleft}stretch{\textquoteright} scenario. $E_{AT},E_{GC}$ - calculated  opening energies  as the difference between the Watson-Crick configuration (minimum) and configuration when central hydrogen bond ($N_1$ {\textbullet}{\textbullet}{\textbullet} $N_3$ distance) reaches $3.7$ {\AA} (red points), $\Delta E= E_{GC}-E_{AT}$.}
\label{fig:enerdepen}       
\end{center}
\end{figure}

\begin{table}
\begin{center}
\noindent\caption{Parameters for A{\textperiodcentered}T and G{\textperiodcentered}C Watson-Crick pairs. The hydrogen bond
distance values and $R$ distances are taken from~\cite{Saenger}. The angles
$\alpha_1$, $\alpha_2 $ and binding energy $E$ are calculated in this
paper.}\vskip3mm\tabcolsep4.5pt
\label{tab:param}
\noindent{\footnotesize\begin{tabular}{|c|c|c|c|c|c|c|c|}
\hline
 & $N_6H$ {\textbullet}{\textbullet}{\textbullet} $O_4$ (A{\textperiodcentered}T)

   & $N_1$ {\textbullet}{\textbullet}{\textbullet} $HN_3$ (A{\textperiodcentered}T)

     & $C_2H$ {\textbullet}{\textbullet}{\textbullet} $O_2$ (A{\textperiodcentered}T)

            &  &  &  &  \\
 & $O_6$ {\textbullet}{\textbullet}{\textbullet} $HN_4$ (G{\textperiodcentered}C), &
 $N_1H$ {\textbullet}{\textbullet}{\textbullet} $N_3$ (G{\textperiodcentered}C), &
 $N_2H$ {\textbullet}{\textbullet}{\textbullet} $O_2$ (G{\textperiodcentered}C), &  &  &  &   \\

 & {\AA} & {\AA} & {\AA} & \raisebox{2.5ex}[0cm][0cm]{$R,$ {\AA}}  & \raisebox{2.5ex}[0cm][0cm]{$\alpha_1, grad$} & \raisebox{2.5ex}[0cm][0cm]{$\alpha_2, grad$} & \raisebox{2.5ex}[0cm][0cm]{$E, kcal/mol$} \\
\hline
A{\textperiodcentered}T & 3.0 & 2.9 & 3.4 & 10.44 & 53.7 & 59.0 & -4.72 \\
\hline
G{\textperiodcentered}C & 2.9 & 3.0 & 2.9 & 10.72 & 58.4 & 56.5 & -7.07 \\
\hline
\end{tabular}}
\end{center}
\end{table}
Opening energy ($E_{AT},E_{GC}$) is considered
to be the difference between the energy of Watson-Crick
configuration (energy minimum) and the configuration when central hydrogen
bond is broken (Fig.\ref{fig:enerdepen}, red points). It should be noted that energy estimates in the present work are the same order of magnitude that obtained in the work~\cite[fig.3, curve 2]{poltevShul1984}.
\begin{table}
\begin{center}
\noindent\caption{Estimated in the present work distances and opening energies for pairs which are opening according to two scenarios: {\textquoteleft}stretch{\textquoteright} and {\textquoteleft}stretch after opening{\textquoteright} .  }\vskip3mm\tabcolsep4.5pt
\label{tab:calculations}
\noindent{\footnotesize\begin{tabular}{|c|c|c|c|c|c|}
\hline
 & $N_6H$ {\textbullet}{\textbullet}{\textbullet} $O_4$ (A{\textperiodcentered}T)

   & $N_1$ {\textbullet}{\textbullet}{\textbullet} $HN_3$ (A{\textperiodcentered}T)

     & $C_2H$ {\textbullet}{\textbullet}{\textbullet} $O_2$ (A{\textperiodcentered}T)

            &  &     \\
 & $O_6$ {\textbullet}{\textbullet}{\textbullet} $HN_4$ (G{\textperiodcentered}C), &
 $N_1H$ {\textbullet}{\textbullet}{\textbullet} $N_3$ (G{\textperiodcentered}C), &
 $N_2H$ {\textbullet}{\textbullet}{\textbullet} $O_2$ (G{\textperiodcentered}C), &  &     \\

 & {\AA} & {\AA} & {\AA} & \raisebox{2.5ex}[0cm][0cm]{$R,$ {\AA}}  & \raisebox{2.5ex}[0cm][0cm]{$ E, kcal/mol$}   \\
\hline
\multicolumn{6}{|c|}{ {\textquoteleft}stretch{\textquoteright} } \\
\hline
A{\textperiodcentered}T & 3.7 & 3.6 & 4.2 & 0.8 & 4.0 \\
\hline
G{\textperiodcentered}C & 3.7  & 3.7 & 3.7 & 0.8 & 6.4 \\
\hline
\multicolumn{6}{|c|}{ {\textquoteleft}stretch after opening{\textquoteright} } \\
\hline
A{\textperiodcentered}T & 4.2 & 3.7 & 4.04 & 0.5 & 1.7 \\
\hline
G{\textperiodcentered}C & 4.1 & 3.7 & 3.7 & 0.4 & 2.0 \\
\hline
\end{tabular}}
\end{center}
\end{table}

As the result of the calculation (Tabl.\ref{tab:calculations}), the A{\textperiodcentered}T pair opening energy for {\textquoteleft}stretch{\textquoteright} scenario is ${ E}_{\mathit{str}}^{\mathit{AT}}\approx 4.0 kcal/mol$, and for {\textquoteleft}stretch after opening{\textquoteright}  ${E}_{\mathit{so}}^{\mathit{AT}}{\approx}1.7 kcal/mol$ . At the same time for the G{\textperiodcentered}C pair corresponding value for the {\textquoteleft}stretch{\textquoteright} scenario is  ${E}_{\mathit{str}}^{\mathit{GC}}{\approx} 6.4 kcal/mol$, and for the {\textquoteleft}stretch after opening{\textquoteright}  ${E}_{\mathit{so}}^{\mathit{GC}}{\approx}2.0 kcal/mol$ . That is to open a pair according to the {\textquoteleft}stretch{\textquoteright} scenario, it is required about 2-3 times more energy than to open it by
{\textquoteleft}stretch after opening{\textquoteright} scenario. Also our estimates show that the opening of the G{\textperiodcentered}C pair by the {\textquoteleft}stretch{\textquoteright} scenario needs  $\Delta E_{\mathit{str}}{\approx}2.4 kcal/mol$ more energy (see. Tabl.\ref{tab:calculations}) than opening of the A{\textperiodcentered}T pair. At the same time, in the {\textquoteleft}stretch after opening{\textquoteright} scenario this difference is reduced to $\Delta E_{\mathit{so}}{\approx}0.3 kcal/mol$. So the pair opening by the {\textquoteleft}stretch after opening{\textquoteright} scenario is much more energetically favorable than by the {\textquoteleft}stretch{\textquoteright} scenario. And heterogeneity plays a much less important role in the {\textquoteleft}stretch after opening{\textquoteright} scenario, than in the {\textquoteleft}stretch{\textquoteright} scenario. So in the {\textquoteleft}stretch after opening{\textquoteright} scenario the process of consequent base pairs opening can flow cooperatively.

Now let us find out how our estimated energies fit to the experimental data (Fig.\ref{fig:expDepen}). Let us start with the {\textquoteleft}stretch{\textquoteright} scenario. In the experiment (Figs.\ref{fig:expSheme},\ref{fig:expDepen}) energy of the double helix opening is the work of external forces to pull the ball to some distance $l$. And the opening force can be calculated by formula (\ref{for:op_force}). Opening energy of the double helix is the sum of the pair opening energy ($E_{pair}$) and the elastic energy of two single strands ($2E_{el}$)~\cite{3Bockelmann}. In the paper~\cite{3Bockelmann} it was estimated the total value of single strands extension for the average force of ${\approx} 13 pN$, which is $l = 9.5 {\AA}$. Since  ${E}_{\mathit{elast}}$ for the opening of poly-A{\textperiodcentered}T and poly-G{\textperiodcentered}C chains are approximately the same, we have ${\Delta F}_{\mathit{str},\mathit{so}}=\mathit{\Delta E}_{\mathit{str},\mathit{so}}/l$. So we have a value  ${\Delta F}_{\mathit{str}}{\approx}$16 pN, and ${\Delta F}_{\mathit{so}}$ ${\approx}2 pN$. As it was mentioned in Sec.\ref{exp_an}, the experimental force difference between two plateaus on the force-distance curve, which corresponds to the difference between A{\textperiodcentered}T and G{\textperiodcentered}C opening forces, has a value ${\Delta F}_{\exp }{\approx}1 pN$ (Tabl.\ref{tab:finaltab}). Apparently, the value is much closer to the value $\Delta F$, calculated for DNA melting process, while the value  ${\mathit{\Delta F}}_{\mathit{so}}$ is close to the observed value in the experiment~\cite{1Bockelmann}. It suggests that base pair opening in the unzipping of $\lambda-$phage DNA experiment at low pulling speeds~\cite{1Bockelmann} can take place not according to the {\textquoteleft}stretch{\textquoteright} scenario, but according to the {\textquoteleft}stretch after opening{\textquoteright} scenario.

\begin{table}
\begin{center}
\noindent\caption{Our calculations of the value $\Delta E= E_{GC}-E_{AT}$, corresponding force differences ${\Delta}F$  measured by the method described in~\cite{3Bockelmann} and the corresponding experimental difference ${\Delta}F_{exp}$. $^*$estimated from DNA melting experiments (Tabl.\ref{tab:force_exp}), $^{**}$estimated from the experiment~\cite{1Bockelmann}.}\vskip3mm\tabcolsep4.5pt
\label{tab:finaltab}
\noindent{\footnotesize\begin{tabular}{|c|c|c|c|}
\hline
  & $\Delta E, kcal/mol$ & $\Delta F, pN$ & $\Delta F_{exp}, pN$ \\
  \hline
 {\textquoteleft}stretch{\textquoteright} & 2.4 & 16 & 7-12$^{*}$ \\
\hline
 {\textquoteleft}stretch after opening{\textquoteright} & 0.3 & 2 & 1$^{**}$ \\
\hline
\end{tabular}}
\end{center}
\end{table}

\section{Discussion and Conclusions}
\label{discuss}
According to our estimations the mechanism which is proposed here describes the DNA double helix unzipping process that can take place according to two different scenarios. 

The first from them ({\textquoteleft}stretch after opening{\textquoteright}) is connected with the base pair transition into the {\textquoteleft}pre-opened{\textquoteright} metastable state along the {\textquoteleft}opening{\textquoteright} pathway (Fig.\ref{fig:scenario}b), the second is the direct extension of the hydrogen bonds along the {\textquoteleft}stretch{\textquoteright} pathway (Fig.\ref{fig:scenario}a).

As the frequency of the bases vibrations in a pair along the {\textquoteleft}opening{\textquoteright} pathway (${\sim}20cm^{-1}$) is lower than along the {\textquoteleft}stretch{\textquoteright} pathway (${\sim}100cm^{-1}$)~\cite{VolkovKosevich1991}, in the first scenario bases in a pair need more time to separate and the first scenario is more probable to take place  at low pulling velocities. It should be noted that the transition into metastable state can occur because of the structural changes of the DNA site situated between the unzipping fork and the dsDNA coil (Fig.\ref{fig:expSheme}). As the result of our estimations it is shown that the experimental difference between two plateaus on the force-distance curve of the $\lambda$-DNA unzipping with low pulling velocities is connected with the realization of the {\textquoteleft}stretch after opening{\textquoteright} scenario.

 In the works~\cite{VolkovKryachko, volkovUnz2009} it is shown that the {\textquoteleft}pre-opened{\textquoteright} metastable state can be formed with the participation of water molecules (Fig.\ref{fig:ATpair}b). It should be noted that our estimations for the difference in opening energies between A{\textperiodcentered}T and  G{\textperiodcentered}C pairs for {\textquoteleft}stretch{\textquoteright} and {\textquoteleft}stretch after opening{\textquoteright} scenarios ( ${\Delta E}_{\mathit{str}}$ and  ${\Delta E}_{\mathit{so}}$) are also appropriate for transition from the state where water is integrated to the external hydrogen bond ($N_{6}H${\textbullet}{\textbullet}{\textbullet}$O_{4}$ for A{\textperiodcentered}T (Fig.\ref{fig:ATpair}) and $O_{6}${\textbullet}{\textbullet}{\textbullet}$HN_{4}$ for G{\textperiodcentered}C)~\cite[fig.4]{VolkovKryachko}. This is because the contribution from interaction of atoms forming the outer hydrogen bond with a water molecule is the same for A{\textperiodcentered}T and G{\textperiodcentered}C base pairs. Further the process of base pairs opening along the {\textquoteleft}stretch{\textquoteright} pathway can take place cooperatively. Cooperativity can emerge because the energy difference for further opening between A{\textperiodcentered}T and G{\textperiodcentered}C pairs (${\Delta E}_{\mathit{so}}$) is very smal (Tabl.\ref{tab:finaltab}, {\textquoteleft}stretch after opening{\textquoteright}). About the cooperativity of the unzipping process in this scenario argues the smooth descent of peaks on the force-distance curve (Fig.\ref{fig:expDepen}) as well.

It is known that the unzipping fork is being blocked from time to time during the unzipping process that results in the peak occurrence on the force-distance curve~\cite{3Bockelmann}. In the present study it is shown that the reason of its blocking is the absence of the conditions of its propagation along the site situated between the fork and the dsDNA coil. This conditions are created after base pair transition into the {\textquoteleft}pre-opened{\textquoteright} state in this site.

At high pulling speeds (${>}1\mu m/sec$) the process becomes to be non-equilibrium, about what argues the hysteresis occurrence on the force-distance curve~\cite{4Bockelmann}. In this case the rotational friction torque starts to play a significant role~\cite{Destainville2016,4Bockelmann} and pairs do not have time to transit into the metastable state. In this case only the second scenario can take place when base pairs open along the {\textquoteleft}stretch{\textquoteright} pathway which has much higher vibrational frequency than the {\textquoteleft}opening{\textquoteright} pathway. The difference between the two plateaus on a force-distance dependence of the $\lambda$-phage DNA should be much bigger than in the first scenario (Tabl.\ref{tab:finaltab}). The unzipping of short DNA hairpins can also take place according to this scenario because the values of the difference between opening forces of short poly-A{\textperiodcentered}T and poly-G{\textperiodcentered}C hairpins (Sec.\ref{exp_an}) and the $\Delta F_{str}$ (Tabl.\ref{tab:finaltab}) are similar.

 As it is seen from the experiment, in natural conditions the process takes place with small velocities (${\approx}10 bp/s$)~\cite{wang2007}. Thus the {\textquoteleft}stretch after opening{\textquoteright} scenario is more probable to take place during the real \textit{in vivo} unzipping processes  such as transcription and translation.


\begin{thebibliography}{99}
\bibitem {smith1992}   S.B. Smith, L. Finzi, C. Bustamante \textit{Direct mechanical measurements of the elasticity of single DNA molecules by using magnetis beads},  Science \textbf{258}, 1122-1126 (1992).
\bibitem {simmons1996}  R.M. Simmons, J.T. Finer, S. Chu,  J.A. Spudich \textit{Quantitative measurements of force and displacement using an optical trap}, Biophys. J. \textbf{70}, 1813-1822 (1996).
\bibitem {3Bockelmann}  U. Bockelmann, B. Essevaz-Roulet, F. Heslot \textit{DNA strand separation studied by single molecule force measurements}, Phys. Rev. E \textbf{58}(2) 2386-2394 (1998).
\bibitem {lavery2002} R. Lavery, A. Lebrun, J.-F. Allemand, D. Bensimon, V. Croquette \textit{Structure and mechanics of single biomolecules: experiment and simulation}, J. Phys. Condens. Matter \textbf{14}, 383-414 (2002).
\bibitem {bustamante1995} C. Bustamante and D. Keller \textit{Scanning force microscopy in biology}, Phys. Today \textbf{48}(12), 32-38 (1995).
\bibitem {bustamante2000} C. Bustamante, S. B. Smith, J. Liphardt, D. Smith \textit{Single-molecule studies of DNA mechanics}, Curr. Opin. Struct. Biol. \textbf{10}(3), 279-285 (2000).
\bibitem {bustamante2003} C. Bustamante, Z. Bryant, S. B. Smith \textit{Ten years of tension: single-molecule DNA mechanics}, Nature \textbf{421}, 423-427 (2003).
\bibitem {Destainville2016} M. Manghi, , N. Destainville \textit{Physics of base-pairing dynamics in DNA}, Phys. Rep. \textbf{631}, 1–41  (2016).
\bibitem {owczarzy2004} R. Owczarzy, Y. You, B.G. Moreira, J.A. Manthey, L. Huang, M. A. Behlke, J. A. Walder \textit{Effects of Sodium Ions on DNA Duplex Oligomers: Improved Predictions of Melting Temperatures}, Biochemistry \textbf{43}, 3537-3554 (2004).
\bibitem {plrevmfk} M.D. Frank-Kamenetskii, S. Prakash \textit{Fluctuations in the DNA double helix: A critical review}, Phys. Life Rev. \textbf{11}, 153-170 (2014).
\bibitem {wartell1985PhReport} R.M. Wartell, A.S. Benight \textit{Thermal denaturation of DNA molecules: a comparison of theory with experiment}, Phys. Rept. \textbf{126} (2), 67-107 (1985).
\bibitem {Breslauer1986} K.J. Breslauer, R. Frank, H. Blocker, L.A. Marky \textit{Predicting DNA duplex stability from the base sequence}, Proc. Natl. Acad. Sci. USA \textbf{83}, 3746-3750 (1986).
\bibitem {mfk2006} P. Yakovchuk, E. Protozanova, M.D. Frank-Kamenetskii \textit{Base-stacking and base-pairing contributions into thermal stability of the DNA double helix}, Nucl. Acids Res. \textbf{34}(2), 564-574 (2006).
\bibitem {rief1999} M. Rief, H. Clausen-Schaumann, H.E. Gaub, \textit{Sequence-dependent mechanics of single DNA molecules}, Nat. Struct. Biol. \textbf{6},  346-349 (1999).
\bibitem {1Bockelmann} B. Essevaz-Roulet, U. Bockelmann, F. Heslot \textit{Mechanical separation of the complementary strands of DNA}, Proc. Natl. Acad. Sci. USA \textbf{94}, 11935-11940 (1997).
\bibitem {2Bockelmann} U. Bockelmann, B. Essevaz-Roulet, F. Heslot \textit{Molecular Stick-Slip Motion Revealed by Opening DNA with Piconewton Forces}, Phys. Rev. Lett. \textbf{79}(22), 4489-4492 (1998).
\bibitem {4Bockelmann} Ph. Thomen, U. Bockelmann, F. Heslot \textit{Rotational Drag on DNA: A Single Molecule Experiment}, Phys. Rev. Lett. \textbf{88}(24), 248102 (2002).
\bibitem {5Bockelmann} U. Bockelmann, Ph. Thomen, B. Essevaz-Roulet, V. Viasnoff, F. Heslot\textit{ Unzipping DNA with Optical Tweezers: High Sequence Sensitivity and Force Flips}, Biophys. J. \textbf{82}, 1537-1553 (2002).
\bibitem {6Bockelmann} U. Bockelmann, P. Thomen, and F. Heslot \textit{Dynamics of the DNA Duplex Formation Studied by Single Molecule Force Measurements}, Biophys. J. \textbf{87}, 3388-3396 (2004).
\bibitem {7Bockelmann} U. Bockelmann, V. Viasnoff \textit{Theoretical Study of Sequence-Dependent Nanopore Unzipping of DNA}, Biophys. J. \textbf{94}, 2716-2724 (2008).
\bibitem {ritort2010} J.M. Huguet, C.V. Bizarro, N. Forns, S.B. Smith, C. Bustamante, F. Ritort \textit{Single-molecule derivation of salt dependent base-pair free energies in DNA},  Proc. Natl. Acad. Sci. USA \textbf{35}, 15431-15436 (2010).
\bibitem {voulgarakis2006} N.K. Voulgarakis, A. Redondo, A.R. Bishop, K.Ø. Rasmussen \textit{Probing the Mechanical Unzipping of DNA}, Phys. Rev. Lett. \textbf{96}, 248101 (2006).
\bibitem {volkovUnz2009} S.N. Volkov, A.V. Solov’yov, \textit{The mechanism of DNA mechanical unzipping}, Eur. Phys. J. D \textbf{54}, 657-666 (2009).
\bibitem {peyrard1989} M. Peyrard, A. Bishop \textit{Statistical mechanics of a nonlinear model for DNA denaturation}, Phys. Rev. Lett. \textbf{62}, 2755-2758 (1989).
\bibitem {Danilovicz} C. Danilowicz, Y. Kafri, R.S. Conroy, V.W. Coljee, J. Weeks, M. Prentiss \textit{Measurement of the Phase Diagram of DNA Unzipping in the Temperature-Force Plane}, Phys. Rev. Lett. \textbf{93}, 078101 (2004).
\bibitem {santalucia1998} J. SantaLucia \textit{A unified view of polymer, dumbbell, and oligonucleotide DNA nearest-neighbor thermodynamics},  Proc Natl Acad Sci USA \textbf{95}(4), 1460-1465 (1998).
\bibitem {liphardt2001} J. Liphardt, B. Onoa, S.B. Smith, I. Tinoco Jr., C. Bustamante \textit{Reversible unfolding of single RNA molecules by mechanical force}, Science \textbf{292}, 733-737 (2001).
\bibitem {levinthal1956} C. Levinthal, H. Crane \textit{On the unvinding of DNA},  Proc. Natl. Acad. Sci. USA \textbf{42} 436-438 (1956).
\bibitem {VolkovMolDynSym} S.N. Volkov, E.V. Paramonova, A.V. Yakubovich, A.V. Solov’yov \textit{Micromechanics of base pair unzipping in the DNA duplex}, J. Phys.: Condens. Matter \textbf{24}, 1-6 (2012).
\bibitem {Saenger}  M. Egli, W. Saenger \textit{Principles of nucleic acids structure}, Springer (1984).
\bibitem {slezkin} N.A. Slezkin \textit{Dynamika viazkoj nezhimaemoj zhidkosti}, Moskov 184-186 (1955).
\bibitem {gross2011unpeeling} P. Gross, N. Laurens, L.B. Oddershede, U. Bockelmann, E.J.G. Peterman, G.J.L. Wuite \textit{Quantifying how DNA stretches, melts and changes twist under tension}, Nat. Phys. \textbf{7}, 731-736 (2011).
\bibitem {nomenclatureEMBO} R.E. Diekmann \textit{Definitions and nomenclature of nucleic acid structure parameters}, EMBO \textbf{8}(1), 1-4 (1989).
\bibitem{VolkovKryachko} E.S. Kryachko, S.N. Volkov \textit{Preopening of the DNA Base Pairs}, Int. J. Quantum Chem. \textbf{82}(4), 193-204 (2001).
\bibitem {VolkovSolovyov2007} S.N. Volkov, A.V. Solov’yov \textit{To the understanding of phase diagram of DNA double helix unzipping}, Biophys. Bull. \textbf{19}(2), 5-12 (2007).
  
\bibitem {VolkovKosevich1991} S.N. Volkov, A.M. Kosevich \textit{Theory of Low-Frequency Vibrations in DNA Macromolecules}, J. Biomolec. Struct. Dynamics \textbf{8}, 1069-1083 (1991).  
  
\bibitem {VolkovMolBiol1995} S.N. Volkov \textit{Priodkrytoye sostoyaniye dvoynoy spirali DNK}, Mol. Biol. \textbf{29}(5), 1086-1094 (1995).
\bibitem {zefirov1974} U.V. Zefirov, P.M. Zorkyi \textit{Van-der-Vaalsovy radiusy atomov v krystallokhimii i strukturnoy khimii}, Zhurnal strukturnoy khimii \textbf{15}, 118-122 (1974).
\bibitem {poltevShul1984} V.I. Poltev, N.V. Shulyupina \textit{Simulation of interactions between nucleic acid bases by refined atom-atom potential functions}, Journal of Biomolecular Structure and Dynamics \textbf{3}(4), 739-765 (1986).
\bibitem {ritort2006} F. Ritort \textit{Single-molecule experiments in biological physics: methods and applications}, J. Phys.: Condens. Matter \textbf{18}, 531-583 (2006).
\bibitem {wang2007} D. S. Johnson, L. Bai, B.Y. Smith, S. S. Patel, M. D.Wang \textit{Single-Molecule Studies Reveal Dynamics of DNA Unwinding by the Ring-Shaped T7 Helicase}, Cell \textbf{129}, 1299-1309 (2007).
\end{thebibliography}
\end{document}